\documentclass[
reprint,
aps,
superscriptaddress,
frontmatterverbose,
amsmath,amssymb,
pra,
]{revtex4-2}

\usepackage{amsfonts}
\usepackage{amsmath}
\usepackage{amssymb}
\usepackage{amsthm}
\usepackage[normalem]{ulem}
\usepackage{graphicx}
\usepackage{dcolumn}
\usepackage{bm}
\usepackage{listings}
\usepackage{xcolor}
\usepackage{csquotes}
\usepackage{hyperref}
\usepackage{array}
\usepackage{booktabs}
\usepackage{multirow}
\usepackage{braket}

\definecolor{codegreen}{rgb}{0,0.6,0}
\definecolor{codegray}{rgb}{0.5,0.5,0.5}
\definecolor{codepurple}{rgb}{0.58,0,0.82}
\definecolor{backcolour}{rgb}{0.95,0.95,0.92}

\lstdefinestyle{mystyle}{
    backgroundcolor=\color{backcolour},
    commentstyle=\color{codegreen},
    keywordstyle=\color{magenta},
    numberstyle=\tiny\color{codegray},
    stringstyle=\color{codepurple},
    basicstyle=\ttfamily\footnotesize,
    breakatwhitespace=false,
    breaklines=true,
    captionpos=b,
    keepspaces=true,
    numbers=left,
    numbersep=5pt,
    showspaces=false,
    showstringspaces=false,
    showtabs=false,
    tabsize=2
}
\hypersetup{
    pdfborder={0 0 0},
    pdfnewwindow=true,
    colorlinks=true,
    hypertexnames=false,
    linkcolor=blue,
    citecolor=blue,
    filecolor=blue,
    urlcolor=blue,
}

\renewcommand\boldsymbol{\bm}

\begin{document}

\title{Accelerating Atom Simulations with Variable-Block Sparse Matrix Library}

\author{Zhanghao Zhouyin}
\affiliation{Department of Physics, McGill University, Montreal, Quebec, Canada H3A 2T8}

\author{Hong Guo}
\affiliation{Department of Physics, McGill University, Montreal, Quebec, Canada H3A 2T8}

\date{\today}

\begin{abstract}
Modern atomistic simulations increasingly employ localized orbitals to represent quantum operators, yielding sparse block matrices whose block shapes vary with chemical species and basis choice. Conventional scalar sparse formats store the entries of each block individually, obscuring this local structure and limiting the use of efficient block algorithms. We present VBCSR, a distributed sparse matrix library that preserves variable-size atomic blocks and accelerates the core linear algebra of large-scale atomistic simulations. A unified interface automatically maps scalar, uniform-basis, and multispecies operators to compressed sparse row (CSR), block sparse row (BSR), or variable-block compressed sparse row (VBCSR). Our advanced acceleration method groups blocks of equal shape and dispatches them to optimized dense kernels. In the reported benchmarks, VBCSR outperforms the tested Python-accessible reference implementations for several block-sparse benchmarks. We further demonstrate VBCSR in an InP nanoparticle application containing more than \(10^6\) atoms.
\end{abstract}

\maketitle

% \section*{Program Summary}

% \begin{description}

% \item[Program title:] VBCSR

% \item[Developer's repository link:]
% https://github.com/DeepElectron/vbcsr

% \item[Licensing provisions:]
% [AGPL-3.0]

% \item[Programming language:]
% C++, Python

% \item[Nature of problem:]
% Localized-orbital atomistic simulations produce sparse
% operators whose natural dense block dimensions depend on
% the chemical species and basis set. Conventional scalar sparse
% formats obscure this structure, while fixed-block representations
% require padding in heterogeneous systems. Efficient distributed
% storage and manipulation of these operators is required for
% large-scale atomistic simulation.

% \item[Solution method:]
% VBCSR represents atomistic operators using a distributed
% atom-centred graph and automatically selects CSR, BSR, or
% variable-block CSR storage according to the local degrees of
% freedom. Blocks with identical dimensions are grouped and
% dispatched to optimized SIMD or BLAS kernels, while MPI,
% OpenMP, and graph-derived communication schedules provide
% distributed execution.

% \end{description}

\section{Introduction}
Localized-orbital methods provide an efficient and physically transparent description of atomistic systems. By associating basis functions with individual atoms, they represent short-range interactions through sparse Hamiltonian, overlap, density, and response operators. This locality provides a compact representation for large-scale systems such as interfaces, defects, disordered materials, moir\'e structures, and multicomponent junctions. As the simulated structure grows, the cost of these calculations is increasingly determined by how these sparse operators are stored, applied, transformed, and distributed across a parallel machine.

The sparse structure of an atomistic operator is naturally block-valued. If atom \(i\) carries \(b_i\) localized orbitals and atom \(j\) carries \(b_j\), their interaction is a dense block of size \(b_i\times b_j\). In a single-species system, these blocks may have a uniform shape, whereas multispecies systems generally produce several shapes because their constituent atoms use different orbital counts. Conventional scalar compressed sparse row (CSR) storage expands these couplings into individual entries, increasing structural metadata and hiding reusable dense work. A uniform-block representation avoids scalar expansion but must pad smaller interactions to the largest block size, introducing values that must be stored, processed, and communicated even though they are absent from the physical model. These representation costs grow with system size and can restrict the scale that can be reached.

Existing electronic-structure packages and high-performance sparse libraries address important parts of this problem. RESCU and CP2K provide scalable capabilities within comprehensive atomistic workflows \cite{michaud2016rescu,kuhne2020cp2k}. DBCSR, NTPoly, BML, and CheSS provide optimized block operations or matrix algorithms for demanding electronic-structure calculations \cite{sivkov2019dbcsr,dawson2018massively,bock2018basic,mohr2017efficient}, while specialized microkernel systems accelerate dense computations within block-sparse workloads \cite{heinecke2016libxsmm}. A complementary need remains for developers of atomistic methods: a lightweight distributed operator layer that preserves species-dependent atomic blocks, scales to large simulations, and can be embedded easily in specialized scientific workflows. Without such a layer, each application must devote substantial effort to managing quantum-operator matrices rather than to implementing the physical model of interest.

\begin{figure*}[t]
    \centering
    \includegraphics[width=0.98\linewidth]{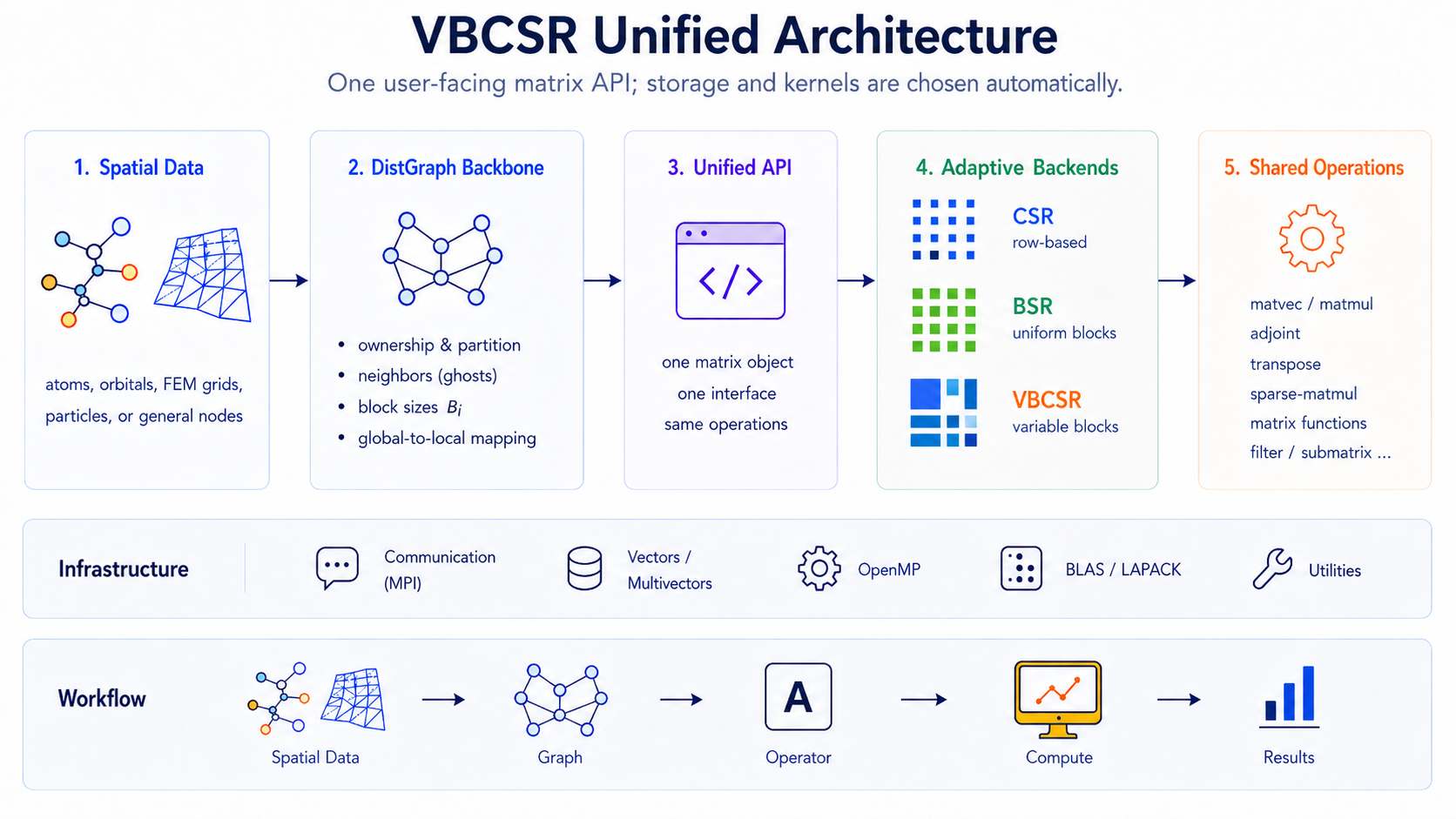}
    \caption{VBCSR uses a shared distributed graph and a unified Python-facing matrix interface across automatically selected CSR, BSR, and VBCSR backends. The backends support common sparse operations through MPI communication, thread-level parallelism, and dense numerical kernels.}
    \label{fig:architecture}
\end{figure*}

We address this need with VBCSR, a distributed sparse matrix library that supports variable block sizes and provides customized high-performance kernels. Each graph node represents an atom and its localized degrees of freedom, while each graph edge stores an atom-atom interaction block. The graph determines the matrix structure, rank ownership, and ghost communication. Blocks of equal shape are stored together, converting irregular atomic connectivity into regular work for optimized dense kernels. To use kernels suited to each structure, VBCSR automatically selects CSR for scalar operators, block sparse row (BSR) for a uniform basis, or variable-block compressed sparse row (VBCSR) for a species-dependent basis. Figure~\ref{fig:architecture} summarizes how VBCSR maps the physical atomic structure to distributed execution.

We first evaluate single-node kernel efficiency and then assess distributed strong and weak scaling. Across scalar, fixed-block, and variable-block operators, VBCSR is competitive with or faster than the tested reference implementations. In particular, variable-block sparse matrix-matrix multiplication is accelerated by up to \(8.4\times\) with one thread and \(3.8\times\) with 16 threads. The distributed experiments then quantify how this performance changes as the number of MPI ranks increases. As an application illustration, we simulate InP nanoparticles using a tight-binding Hamiltonian. We compute the density of states and extract the size-dependent band gap for systems containing up to 1,063,609 atoms, demonstrating the million-atom scalability of VBCSR for large atomistic operators.

\section{VBCSR for Atomistic Simulation}
\subsection{From Localized Atomic Orbitals to Block-Sparse Operators}

Localized atomic-orbital models naturally produce sparse operators with physically meaningful blocks of variable size. Each atom interacts primarily with atoms in its local environment; depending on the model, longer-range interactions either decay with distance or are truncated beyond a prescribed cutoff. The resulting operators are therefore sparse in real space. Each interaction block couples the orbitals carried by two atoms. When every atom carries one degree of freedom, each interaction is a scalar and compressed sparse row (CSR) storage is appropriate. When all atoms carry the same number of orbitals, the interactions form equally sized dense blocks suited to block sparse row (BSR) storage. Multispecies systems are generally heterogeneous: if atom $i$ carries $b_i$ orbitals and atom $j$ carries $b_j$ orbitals, their coupling is a dense $b_i\times b_j$ block. These varying shapes produce the variable-block compressed sparse row (VBCSR) structure. Retaining the blocks preserves the correspondence between numerical data and the atomic model, avoids expanding each coupling into unrelated scalar entries, and avoids padding every coupling to the largest orbital count. Numerical kernels can consequently operate on the compact units generated directly by the atomistic Hamiltonian or overlap operator.

\subsection{Three Acceleration Mechanisms}

VBCSR exploits this atom-centred structure through three complementary acceleration mechanisms. The first is automatic backend selection. The orbital counts recorded by the atomic graph determine whether the matrix follows the CSR, BSR, or VBCSR path, so scalar, uniform-basis, and species-heterogeneous operators each use storage and kernels suited to their structure. This selection gives us one logical matrix abstraction while avoiding the cost of forcing every system through a general variable-block representation. The second mechanism regularizes the heterogeneous case without changing its physical blocks. VBCSR groups atom-atom blocks with equal row and column dimensions and stores them in contiguous pages. A kernel can then process a sequence of identically shaped blocks using single-instruction, multiple-data (SIMD) operations. A routing layer dispatches each dense block operation to either a specialized native SIMD kernel or a Basic Linear Algebra Subprograms (BLAS) routine, according to its shape. Thus, irregularity remains in the atomic graph for compact storage and semantic consistency, while each numerical batch presents regular work to the processor. The third mechanism coordinates data placement with parallel execution. Graph-derived system partitions and exchange schedules are reused across operator applications, while OpenMP in each partition assigns batches of blocks for threaded parallelization. The same partitions guide first-touch allocation on non-uniform memory access (NUMA) systems, improving locality and thereby reducing thread overhead for memory movement. Together, these choices reduce avoidable representation changes, expose regular dense work, and keep computation close to the data.

\subsection{Python Simplicity, Optimized Execution}

The optimized execution model remains accessible through a single Python matrix object. Users can construct and assemble atomistic operators, apply them to vectors or batches of vectors, and combine them through familiar arithmetic. Matrix application and sparse matrix multiplication use the \texttt{@} operator, while scaling, addition, subtraction, transposition, filtering, and matrix-function utilities are exposed through standard operators or matrix methods. In serial mode, compatibility with the SciPy \texttt{LinearOperator} interface allows VBCSR matrices to be used by SciPy solvers. Conversion paths to and from NumPy arrays and SciPy sparse matrices support integration with existing atomistic codes with minimal effort. A single Python expression can therefore initiate ghost synchronization, select the active sparse backend, and execute threaded block kernels without requiring the user to coordinate these steps explicitly. The Python layer hides this complexity, keeping atomistic workflows concise, clean, and purely operator-level while using specialized compiled execution.

\section{Architecture and Algorithms}

\subsection{Data Model and Distributed Ownership}

The distributed graph defines the structure and system partition throughout VBCSR. Each graph node $i$ has a block size $b_i$, which represents its number of scalar degrees of freedom. An MPI rank owns a set of graph nodes and stores the corresponding block rows. Remote nodes connected by those rows are represented locally as ghosts. Local indices place all owned nodes before all ghosts, while ghosts are ordered first by owner rank and then by global node identifier. This ordering makes ownership explicit and communication buffers rank-contiguous. Graph connectivity is defined above the block level, while scalar storage and communication ranges can be derived from it.

The graph also constructs reusable communication schedules from this ownership model. The block send and receive indices identify which owned nodes must be exported and which ghost nodes must be requested. These communication patterns are reused by matrix operators, \texttt{DistVector}, and \texttt{DistMultiVector}. For example, standard matrix-vector multiplication synchronizes ghost entries according to the node connectivity recorded in the distribution pattern. For adjoint operations, the data flow is reversed: contributions accumulated for ghost columns are returned to their owners and reduced into the corresponding owned entries. Forward synchronization and adjoint reduction therefore use a common graph-derived description of the communication boundary.

\subsection{Structure-Matched Sparse Algorithms}

VBCSR uses \texttt{BlockSpMat} to represent a matrix. It contains one active CSR, BSR, or VBCSR backend selected from the graph block sizes. The VBCSR backend separates graph order from value order when storing heterogeneous blocks. Backend construction first counts every distinct pair $(b_i,b_j)$ and registers it as a shape class. Each heterogeneous block is then mapped to its location in the shape-specific buffer. A logical block can therefore be retrieved through the map, while disconnected blocks with the same shape can be passed directly to batched operations without repacking.

Numerical kernels in VBCSR use a shape-based routing mechanism. This mechanism supports specialized kernels with vectorized execution for small matrices, while larger blocks fall back to linked BLAS routines. This avoids the interface overhead of calling BLAS for small matrices. OpenMP distributes batches among threads. Output matrices and vectors are constructed using the same thread partitions, so each thread initializes and accesses contiguous storage. We additionally control each thread to touch pages physically stored in the NUMA domain of the corresponding core. This design targets modern CPUs, on which NUMA placement and cache locality matter significantly for performance. The following sections describe the core algorithms supported by VBCSR.

\textit{Matrix multiplication}: This category includes sparse matrix-vector products (SpMV), sparse matrix-dense matrix products (SpMM), and their adjoint operations. Each forward operation begins by synchronizing ghost input entries, after which the backend traverses the owned block rows and accumulates $A_{ij}X_j$. The adjoint operation instead evaluates
\begin{equation}
    Y_j \mathrel{+}= A_{ij}^{\dagger}X_i,\qquad j\in\mathcal{J}(i),
    \label{eq:block_apply_adjoint}
\end{equation}
where $\mathcal{J}(i)$ denotes the stored block columns in row $i$. The operation subsequently reduces contributions associated with ghost columns to their owning ranks.

\textit{Thresholded sparse matrix multiplication}: This is organized as distributed symbolic and numeric stages. At the block level, it forms
\begin{equation}
    C_{ij} = \sum_k A_{ik}B_{kj}.
    \label{eq:spgemm}
\end{equation}
The procedure follows the block-thresholding strategy used in distributed block-sparse multiplication \cite{sivkov2019dbcsr}. It first exchanges the remote row metadata and block norms required to enumerate products involving non-local rows of $B$. The symbolic stage then accumulates candidate output columns and estimates each contribution from $\lVert A_{ik}\rVert\lVert B_{kj}\rVert$. For an accumulated block norm,
\begin{equation}
    \lVert C_{ij}\rVert
    \leq
    \sum_k \lVert A_{ik}\rVert\,\lVert B_{kj}\rVert ,
    \label{eq:spgemm_norm_bound}
\end{equation}
so, for a nonnegative block-norm threshold $\varepsilon$, a candidate is safely rejected when the accumulated upper bound is less than $\varepsilon$. Such filtering is applicable when physical locality or matrix decay makes sufficiently small blocks negligible within the accuracy requirements of the calculation \cite{prodan2005nearsightedness,benzi2013decay,benzi2007decay}.

The surviving symbolic pattern determines the result graph and the data required by the numerical stage. Only remote blocks that participate in retained product paths are requested from their owning ranks. After these blocks arrive, the numerical stage groups products by the triple of row, inner, and column dimensions $(b_i,b_k,b_j)$. Every group therefore contains compatible dense products of shape $(b_i\times b_k)(b_k\times b_j)$ and can use the corresponding row-major scalar, SIMD, or BLAS path. Products accumulate into the blocks of the constructed result graph. A final block-norm filter further removes redundant output blocks.

\subsection{Atomistic Interfaces}
\texttt{AtomicData} translates localized atomistic information into this distributed block-graph model. Each atom has a global identifier, atomic type, position, and species-dependent orbital count. The orbital count becomes the node block size. Distributed neighbour construction assigns owned atoms to ranks and records adjacent local or remote atoms. Periodic edges additionally record their integer lattice shifts. After the graph is constructed, ghost atoms receive the features needed for local operator assembly, including their types, orbital counts, positions, and periodic information. This representation simplifies distributed operator construction for localized-basis atomistic simulations.

\texttt{ImageContainer} organizes periodic real-space operators by lattice translation. Edges are grouped according to their shift $\mathbf{R}$, and each distinct translation is associated with a block-sparse image whose rows follow the same atom partition and species-dependent block-size model. The union of these image graphs defines exactly the connectivity of the underlying \texttt{AtomicData}. The blocks of each real-space matrix are assembled according to the translation-resolved matrices and the atom partition. Sampling a reciprocal-space point combines the images into a complex block-sparse operator,
\begin{equation}
    O(\mathbf{k}) =
    \sum_{\mathbf{R}}
    e^{-2\pi\mathrm{i}\,\mathbf{k}\cdot\mathbf{R}}
    O(\mathbf{R}),
    \label{eq:k_sampling}
\end{equation}
where $\mathbf{R}$ is an integer lattice translation and $\mathbf{k}$ is expressed in fractional reciprocal coordinates. Equation~\eqref{eq:k_sampling} is the default \texttt{R} convention. The alternative \texttt{R+tau} convention multiplies block $(i,j)$ by the phase associated with $\mathbf{R}+\boldsymbol{\tau}_j-\boldsymbol{\tau}_i$. These interfaces connect atom types, localized orbital dimensions, periodic neighbours, real-space blocks, and reciprocal-space sampling to the same distributed operator model.

\section{Benchmark Experiments}
The benchmark experiments addressed two questions. The first was whether VBCSR retained or improved kernel efficiency relative to Python-accessible and optimized references for representative sparse operators. The second was how the same operator families scaled when distribution and ghost communication were introduced across Message Passing Interface (MPI) ranks. The measurements used cluster nodes equipped with AMD EPYC 7532 CPUs with 32 cores per node, Open MPI and OpenMP for parallelization, and MKL with customized BLAS kernels. The matrices represented atom-like periodic geometric graphs with finite cutoffs and comparable mean connectivity across three structural domains: scalar compressed sparse row (CSR), fixed-size block sparse row (BSR), and variable-block compressed sparse row (VBCSR). We tested sparse matrix--vector multiplication (SpMV), sparse matrix--dense matrix multiplication (SpMM), and sparse general matrix--matrix multiplication (SpGEMM) in double precision. For the kernel-efficiency and strong-scaling experiments, the graph size in each domain was chosen to target a matrix storage footprint of approximately 500~MiB. Each reported kernel time was the median of seven timed repetitions following three warm-up iterations.

\subsection{Kernel Efficiency}

\begin{figure*}[t]
    \centering
    \includegraphics[width=0.98\linewidth,trim=0 14pt 0 0,clip]{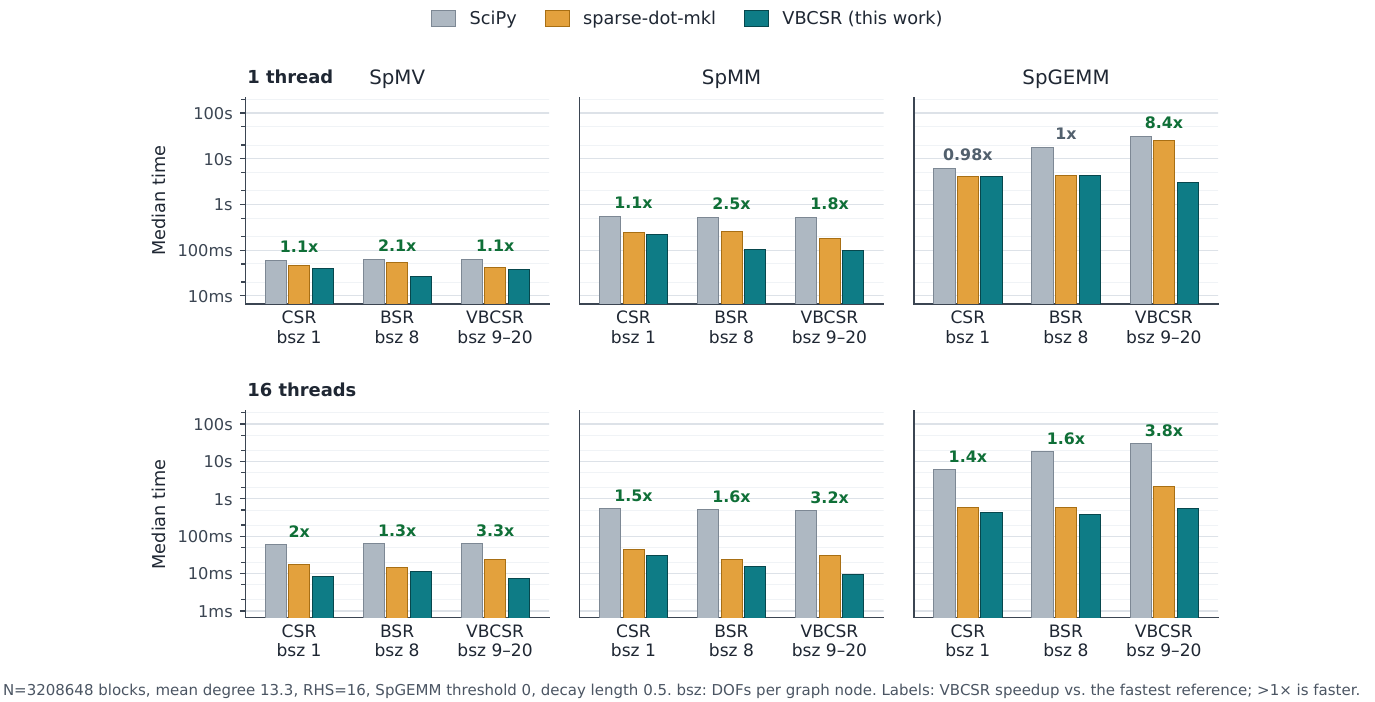}
    \caption{Kernel-efficiency benchmarks for sparse matrix-vector multiplication (SpMV), sparse matrix-dense
    matrix multiplication (SpMM), and sparse general matrix-matrix
    multiplication (SpGEMM).  Each structural domain targets a matrix storage
    budget of approximately 500~MiB; graph block counts therefore differ by domain. CSR uses scalar blocks, BSR uses block size 8, and VBCSR uses
    variable block sizes from 9 to 20. SpMM uses 16 right-hand sides, and the reported SpGEMM case uses zero threshold. Bars report the median kernel time over
    seven timed repetitions following three warm-up iterations.  Speedup labels
    report VBCSR performance relative to the fastest reference implementation;
    values larger than one indicate that VBCSR is faster.}
    \label{fig:benchmark_efficiency}
\end{figure*}

Kernel efficiency was evaluated against SciPy and Intel MKL sparse kernels accessed through \texttt{sparse-dot-mkl}. SciPy represented a standard sparse Python path, while \texttt{sparse-dot-mkl} provided an optimized vendor reference available through Python. The thread count was set to one or 16. In the variable-block domain, the reference libraries operated using CSR matrices because their standard interfaces did not directly represent variable block sizes. Figure~\ref{fig:benchmark_efficiency} reports the median kernel times and the speedup of VBCSR relative to the fastest available reference in each case.

The three operations exposed distinct performance trends within each structural domain. SpMV performed one block multiplication for each stored block and remained relatively memory-bound. SpMM applied the same sparse operator to multiple right-hand sides and therefore allowed greater reuse of matrix values within dense kernels. SpGEMM generated many block products, creating more opportunities to benefit from preserving block structure and batching products of equal shape. These qualitative trends were consistent with the acceleration mechanisms described earlier.

Performance in the scalar CSR domain showed the behaviour of ordinary scalar sparse matrices in VBCSR. The one-thread cases were near parity with the reference implementations. At 16 threads, VBCSR speedups ranged from approximately \(1.4\) to \(2.0\), depending on the operation. These observations indicate that backend dispatch did not impose a large penalty for the tested scalar operators.

Fixed-block BSR exposed more regular dense work than scalar CSR. With block size 8, the one-thread SpMV and SpMM cases were approximately \(2.1\)--\(2.5\) times faster than the references. The corresponding 16-thread cases retained speedups of approximately \(1.3\)--\(1.6\). These measurements are consistent with the fixed-block dense kernels, SIMD-aware batched operations, and NUMA-aware memory placement described earlier.

The variable-block domain produced the largest improvements for operations with greater block-level reuse. For block sizes from 9 to 20, one-thread SpMV remained close to parity with the fastest scalar-expanded reference. Under the same thread count, SpMM and SpGEMM were faster by factors of approximately \(1.8\) and \(8.4\), respectively. With 16 threads, the SpMV, SpMM, and SpGEMM speedups were approximately \(3.3\), \(3.2\), and \(3.8\). The observed advantage beyond SpMV was consistent with our design: preserving heterogeneous blocks, avoiding scalar CSR representation, and grouping compatible products by shape. In particular, SpMM reused operator data across multiple right-hand sides, whereas SpGEMM presented shape-batched block products to the dense kernels. These results remain specific to the tested operators, block-size range, thread counts, and references. Therefore, we report the observed speedups as support for the competitive performance of VBCSR, rather than as evidence of general superiority.

\begin{figure*}[t]
    \centering
\includegraphics[width=0.98\linewidth,trim=0 16pt 0 0,clip]{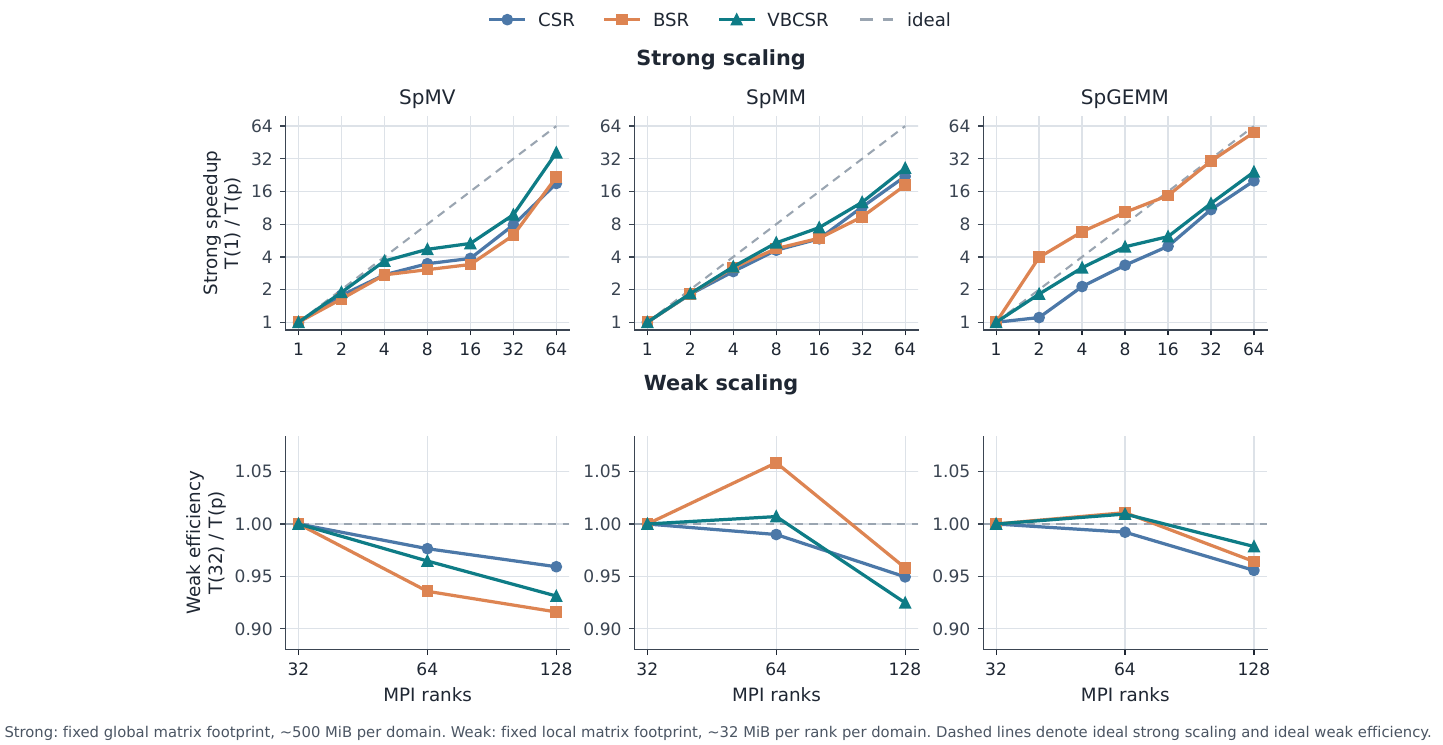}
    \caption{Distributed scaling benchmarks.  Strong scaling uses a fixed
    global matrix storage target of approximately 500~MiB per structural domain. Weak scaling keeps the local matrix size approximately fixed
    at 32~MiB per rank and per domain, and increases the global problem size in proportion to the MPI rank
    count. Strong-scaling speedup is measured as \(S(p)=T(1)/T(p)\), and weak-scaling efficiency is defined as \(E_{\mathrm{w}}(p)=T(32)/T(p)\), where \(T\) denotes the execution time. Dashed curves mark ideal strong scaling and ideal weak-scaling efficiency.}
    \label{fig:benchmark_scaling}
\end{figure*}

\subsection{Distributed Scaling and Reproducibility}

Distributed scaling tested whether the storage and kernel paths remained effective after MPI distribution and ghost communication were included. Strong scaling held the global problem fixed while increasing the number of ranks from 1 to 64. Weak scaling instead held the local matrix footprint approximately fixed and increased the global problem in proportion to the rank count. The former measured the benefit of dividing a finite amount of work, whereas the latter tested whether communication and synchronization costs remained controlled as the distributed graph grew.

Figure~\ref{fig:benchmark_scaling} shows good strong scaling across all three structural domains, with the scaling rate depending on the arithmetic intensity of the operation. Because the global matrix footprint was held constant at approximately 500~MiB, increasing the rank count reduced the local computational workload per rank and made communication, synchronization, and scheduling overheads more prominent. SpGEMM generally scaled further than SpMV and SpMM, consistent with its higher ratio of computation to communication. The curves demonstrate that the implementation benefits from additional MPI ranks over the tested range, while also showing operation-dependent scaling limits.

Weak scaling shows behaviour close to ideal over the tested range. With the local matrix footprint held approximately constant, the measured efficiencies ranged from roughly \(0.92\) to \(1.06\) as the rank count increased from 32 to 128. Equivalently, the total runtime changed by less than \(10\%\) while the global problem size was quadrupled. These observations indicate that the overhead associated with distribution and ghost exchange remained controlled over this range, without evident saturation.

\begin{figure*}[t]
    \centering
\includegraphics[width=0.98\linewidth,trim=0 16pt 0 0,clip]{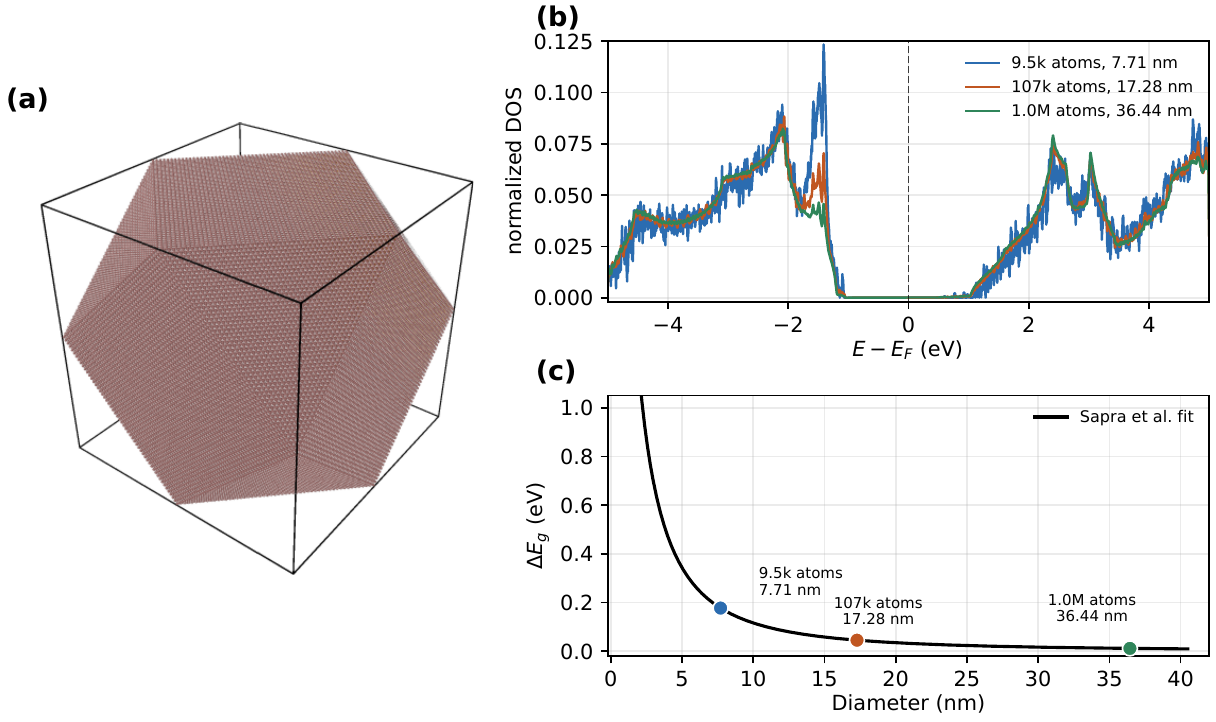}
    \caption{Atomistic density-of-states and band-gap calculation for hydrogen-passivated zinc-blende InP nanoparticles.
(a) Visualization of the largest completed particle, containing 1,063,609 atoms
including surface H passivation, with an effective diameter of 36.44~nm.
(b) Normalized DOS for the three particle sizes studied here:
12,083 atoms ($D=7.71$~nm), 120,250 atoms ($D=17.28$~nm), and
1,063,609 atoms ($D=36.44$~nm). The DOS is computed from the
In($sp^3$)--P($sp^3d^5$)--H($s$) tight-binding Hamiltonian with interactions
through second-nearest neighbours, using the parameterization of
Sapra \textit{et al.}~\cite{sapra2003accurate}. The spectra are aligned to a
common mid-gap reference energy and shown as normalized DOS per orbital.
(c) Diameter dependence of the confinement-induced gap shift, $\Delta E_g$.
The solid curve is the fit reported by Sapra \textit{et al.},
$\Delta E_g = 100/(5.8D^2 + 27.2D + 10.4)$, with $D$ in nm; colored markers
indicate the three particle diameters used in the present DOS calculations.}
    \label{fig:InP}
\end{figure*}

\section{Large-Scale InP Nanoparticle Demonstration}
In this section, we apply VBCSR to a material-specific atomistic calculation. We study finite InP nanoparticles with the zinc-blende crystal structure, for which quantum confinement produces a pronounced size dependence in the electronic gap \cite{sapra2003accurate}. Reaching the bulk-like regime requires particle dimensions comparable to or larger than the relevant excitonic length scale, which is on the order of tens of nanometres for InP \cite{sapra2003accurate}.

Using a reported InP excitonic length scale of approximately 15~nm \cite{sapra2003accurate}, we include particle diameters on both sides of this scale. For direct comparison with prior results, we construct particles with diameters of 7.71, 17.28, and 36.44~nm, containing 12,083, 120,250, and 1,063,609 atoms, respectively. The tight-binding basis comprises In($sp^3$), P($sp^3d^5$), and H($s$) orbitals, and the Hamiltonian includes interactions through second-nearest neighbours, following the selected parameterization \cite{sapra2003accurate}. Each system is surface-passivated by hydrogen atoms and studied using periodic boundary conditions with a vacuum region to isolate it. We use these three systems for the DOS and gap analysis. The largest completed calculation therefore provides a direct demonstration of VBCSR at the million-atom scale.

The reported properties include the density of states (DOS) and the band gap, as displayed in Figure~\ref{fig:InP}. Panel (a) shows the million-atom nanoparticle geometry, and panels (b) and (c) show the DOS and the size-dependent band-gap shifts. The gap shift is computed from the difference between the finite InP nanoparticle band gap estimated from the KPM DOS spectrum and the bulk band gap, as also defined in \cite{sapra2003accurate}. We compute the DOS using the kernel polynomial method (KPM) with 147,456 moments, corresponding to an energy resolution of 0.987~meV under the chosen spectral scaling and damping kernel, and then estimate the gap from the low-DOS interval using a consistent KPM-resolved threshold of $10\%$ of its resolution. The estimated gap shifts are consistent with the reported fit of Sapra \textit{et al.}~\cite{sapra2003accurate}, demonstrating that VBCSR can support physical atomistic simulations with large-scale computational requirements.

\section{Conclusion}

VBCSR provides lightweight, scalable, and efficient distributed infrastructure for localized-basis atomistic simulation. Its atom-centred graph maps each atom to its localized degrees of freedom and each physical interaction to a dense matrix block, while the library manages sparse storage, distribution, communication, and kernel selection. Supporting CSR, BSR, and true variable-block CSR backends through one interface allows single-orbital, uniform-basis, and multispecies systems to use structure-matched data layouts and optimized kernels without changing the application-level workflow. The benchmarks demonstrate the performance of these backends on the tested CPU platform, while the InP application demonstrates the library's capacity for large atomistic systems. The same graph-and-block abstraction may also support variable-space finite-element PDE discretizations, which remains for future exploration.

\section*{Acknowledgments}
Hong Guo acknowledges financial support from the Natural Sciences and Engineering Research Council of Canada (NSERC). Zhanghao Zhouyin thanks the Department of Physics, McGill University, for Dr.\ and Mrs.\ Milton Leong Fellowships in Science. This work benefits from the co-authors' RQMP membership, \url{https://doi.org/10.69777/309032}. We thank the Digital Research Alliance of Canada for substantial computational resource allocations that made this work possible.

\section*{Declaration of generative AI}
During the preparation of this work, the authors used ChatGPT (OpenAI) and Claude (Anthropic) to assist with language editing and improving the clarity and presentation of the manuscript. After using these tools, the authors reviewed and edited the content as needed and take full responsibility for the content of the publication.

\section*{Code and data availability}

The scripts for generating the current benchmark data and the VBCSR library are
openly available at \url{https://github.com/DeepElectron/vbcsr}.  At
publication, the exact code revision, benchmark inputs, and plotting scripts
used for the figures should be archived with the manuscript.

\bibliography{main.bib}
\end{document}